\documentclass[runningheads]{llncs}
\usepackage{graphicx}
\usepackage{mathtools} 
\usepackage{amsfonts}
\usepackage{lmodern,bm}  
\usepackage{booktabs, makecell}
\usepackage[misc,geometry]{ifsym} 
\makeatletter
\newcommand{\printfnsymbol}[1]{%
  \textsuperscript{\@fnsymbol{#1}}%
}
\def\@fnsymbol#1{\ensuremath{\ifcase#1\or *\or \dagger\or \ddagger\or
   \mathsection\or \mathparagraph\or \|\or **\or \dagger\dagger
   \or \ddagger\ddagger \else\@ctrerr\fi}}
\makeatother
\newcommand{\etal}{\textit{et al}. }

\begin{document}
\title{Hierarchical Phenotyping and Graph Modeling of Spatial Architecture in Lymphoid Neoplasms}
\titlerunning{Graph Modeling Cellular Spatial Architecture}
\author{Pingjun Chen\inst{1}\orcidID{0000-0003-0528-1713}\thanks{Equal Contribution} \and Muhammad Aminu\inst{1}\orcidID{0000-0002-9903-8812}\printfnsymbol{1} \and Siba El Hussein\inst{2}\orcidID{0000-0002-1338-7107}\printfnsymbol{1} \and Joseph D. Khoury\inst{3}\orcidID{0000-0003-2621-3584}\thanks{Co-supervision} \and Jia Wu\inst{1}\orcidID{0000-0001-8392-8338}\printfnsymbol{2}\Letter}
\authorrunning{P. Chen \etal}
\institute{Department of Imaging Physics, Division of Diagnostic Imaging, The University of Texas MD Anderson Cancer Center, Houston, TX, USA \\
\email{jwu11@mdanderson.org} 
\and Department of Pathology, University of Rochester Medical Center, Rochester, NY, USA \and Department of Hematopathology, Division of Pathology and Lab Medicine, The University of Texas MD Anderson Cancer Center, Houston, TX, USA}
\maketitle
\begin{abstract}

The cells and their spatial patterns in the tumor microenvironment (TME) play a key role in tumor evolution, and yet the latter remains an understudied topic in computational pathology. This study, to the best of our knowledge, is among the first to hybridize local and global graph methods to profile orchestration and interaction of cellular components. To address the challenge in hematolymphoid cancers, where the cell classes in TME may be unclear, we first implemented cell-level unsupervised learning and identified two new cell subtypes. Local cell graphs or supercells were built for each image by considering the individual cell's geospatial location and classes. Then, we applied supercell level clustering and identified two new cell communities. In the end, we built global graphs to abstract spatial interaction patterns and extract features for disease diagnosis. We evaluate the proposed algorithm on H\&E slides of 60 hematolymphoid neoplasms and further compared it with three cell level graph-based algorithms, including the global cell graph, cluster cell graph, and FLocK. The proposed algorithm achieved a mean diagnosis accuracy of 0.703 with the repeated 5-fold cross-validation scheme. In conclusion, our algorithm shows superior performance over the existing methods and can be potentially applied to other cancer types.

\end{abstract}


\section{Introduction}
The tumor is a complex ecosystem that emerges and evolves under selective pressure from its microenvironment, involving trophic, metabolic, immunological, and therapeutic factors. The relative influence of these factors orchestrates the abundance, localization, and functional orientation of cellular components within the tumor microenvironment (TME) with resultant phenotypic and geospatial variations, a phenomenon known as intratumoral heterogeneity (ITH)~\cite{vitale2021intratumoral,gerlinger2012intratumor}. As such, ITH provides a substrate from which neoplastic cells emerge, escape immunologic surveillance, undergo genetic evolution, and develop pathways that lead to therapy resistance. In routine practice, such attributes are evaluated microscopically primarily using tumor tissue sections stained with the hematoxylin and eosin stain (H\&E). With the advent of digital pathology, machine learning-empowered computational pipelines have been proposed to profile TME using H\&E tissue sections to enhance cancer diagnosis and prognostication~\cite{janowczyk2016deep,komura2018machine,wu2020integrated}.

Most studies published to date have focused on phenotyping the textural patterns of tissue slides in a top-down manner, through either conventional hand-crafted features (Gabor, Fourier, LBP, GLCM, etc.) or deep convolutional neural networks (CNN) to extract versatile features specifically tailored for particular clinical scenarios~\cite{hou2016patch,zhu2017wsisa,campanella2019clinical,li2020rule,chen2020interactive,chen2021automatic,zhang2019text}. Though these studies have achieved promising performance, they missed connection to individual cellular components in TME, thereby limiting their ability to capture neoplastic cell characteristics in their microenvironment milieu and capitalizing on critical diagnostic aspects and prognostic predictions. To bridge this gap, few bottom-up studies focused on profiling cellular architectures from digital pathology slides have emerged using graph theory approach and graph convolution network (GCN) approach~\cite{shin2015quantitative,lewis2014quantitative,lu2021feature,zhou2019cgc,pati2020hact,jaume2020towards}. The graph theory approach first constructs either local or global graph structures and then extracts hand-crafted features to test their clinical relevance. By contrast, the GCN approach aims to automatically learn representations from the global graph formed at the cellular level and abstract the features via multiple layers of graph convolution and pooling operations, similar to the CNN approach. However, a common limitation to these algorithms is their lack of ability to differentiate known cellular populations. Pilot studies have been proposed to address this limitation in the colorectal cancer~\cite{javed2020cellular,schurch2020coordinated}, a solid tumor with a well-studied TME that contains vascular structures, immune cells, fibroblasts, and extracellular matrix~\cite{joyce2015t,korneev2017tlr}. There are no studies exploring this approach in hematolymphoid neoplasms. 

Diffuse large B-cell lymphoma (DLBC) is an aggressive lymphoid malignancy that may arise through a series of clonal evolution steps, occasionally from a low-grade lymphoid neoplasm called chronic lymphocytic leukemia/small lymphocytic lymphoma (CLL/SLL)~\cite{agbay2016high,el2021artificial}. The latter is among the most prevalent lymphoid neoplasms and is associated with a favorable survival rate (5-year survival rate around 83\%). In a subset of patients, CLL/SLL can pursue a more aggressive clinical course referred to as “accelerated CLL/SLL” (aCLL). Patients with CLL/SLL and/or aCLL can also develop a full-blown DLBCL, called Richter transformation DLBCL (RT-DLBL), a complication that is associated with a high mortality rate with a median overall survival of 9 months~\cite{agbay2016histologic,el2021artificial}. Unlike solid tumors with a well-defined TME, there is no standard way to categorize cells into biological subtypes in hematolymphoid tumors. To diagnose these patients, pathologists make diagnoses based on their empirical knowledge of tumor cell morphology, normal histology, and deviations from normal tissue architectures. By default, such an approach is subjective and prone to inter- and intra-reader variations.

\begin{figure}
\includegraphics[width=\textwidth]{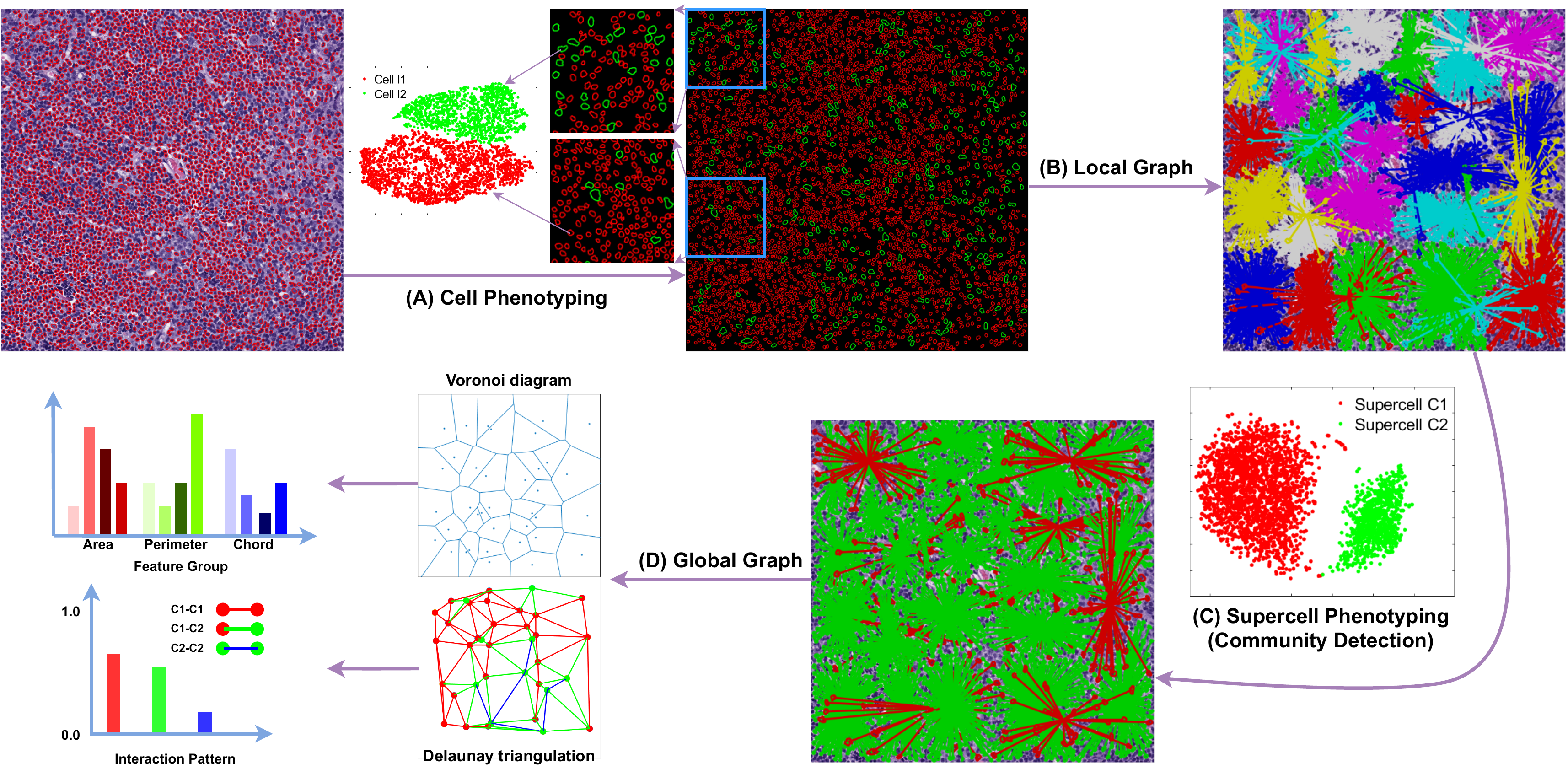}
\caption{Illustration of the proposed hierarchical cell phenotyping and graph modeling algorithm to profile the cellular architecture for analyzing lymphoid pathology images. (A) Cell phenotyping via spectral clustering algorithm. (B) Cell local graph construction to obtain supercells via a Ranking Operation-based Clustering (ROC) algorithm. (C) Supercell phenotyping (community detection) via spectral clustering algorithm. (D) The construction of global graphs based on supercells for extracting features.}
\label{fig:spatial_pattern_pipeline}
\end{figure}

To improve the diagnostic accuracy of tissue examination of lymphoid neoplasms, we propose an innovative computational framework that integrates unsupervised clustering and graph modeling algorithms as a knowledge discovery approach to hierarchically decode cellular and clonal level phenotypes in TME. In particular, we dissect the process into four key steps (Fig.~\ref{fig:spatial_pattern_pipeline}). First, we segment each cell and based on their features to identify intrinsic subtypes. Second, we focus on spatial interaction among neighboring cells by building local graphs factoring in their subtypes, so that closely-interacting cells are merged to form supercells. Third, we pool the supercells together to discover the cellular community at a population level. Lastly, we build global graphs incorporating community information to extract features for diagnostic purposes.

To sum up, the main contributions of our work are as follows: (1) We use clustering as a knowledge discovery approach to explore intrinsic cellular and clonal phenotypes of TME to advance our understanding of hematolymphoid cancers. (2) We develop a novel hierarchical graph construction algorithm to comprehensively characterize interaction patterns at both the local cellular level and global community level for cellular architectures. (3) The proposed algorithm attains superior diagnostic performance compared with three state-of-the-art graph theory-based methods to enhance blood cancer diagnosis. To the best of our knowledge, this is the first study using such a design in the digital hematopathology field. We believe that our design has opened a new direction to quantify intratumoral heterogeneity, an aspect that has particular clinical implications in terms of risk stratification and prediction.

\section{Method}
\subsection{Cell Phenotyping via Unsupervised Learning}
Given a digital H\&E slide, segmentation of nuclei is a prerequisite step for cellular-based analysis. We adopt a multiple-pass adaptive voting (MPAV) algorithm to segment the nuclei in digitized tissue images~\cite{lu2016multi}. To discover cell subtypes, we first characterize the morphology, intensity, and regional patterns of individual nuclei with a set of features (n=24, ``Cell features'' in Table~\ref{tab:cell_features}). To eliminate redundant features, we apply the Laplacian score method to select the top informative ones for the downstream analysis~\cite{he2005laplacian}. By treating each cell as an instance and randomly sampling cells across three disease subtypes, we embed pooled cells in a two-dimensional t-SNE feature space and then apply the spectral clustering to discover inter- and intra-disease cell subtypes. After repeating the sampling process multiple times, we observe that two intrinsic cell phenotypes are consistently identified. We train a classifier via an ensemble of learners aggregated with the bagging method~\cite{friedman2001elements} based on sampled cells for clustering, so that the cell subtypes can be propagated to unsampled ones.

\begin{table}
\centering
\caption{Summary of manual-crafted cell and graph features.}
\label{tab:cell_features}
\begin{tabular}{|c|c|l|}
\hline
Category &  No. & Feature Names \\
\hline
Cell features & 24 & 
            \begin{tabular}{@{}l@{}} 
                \textbf{Area}, \textbf{EquivDiameter}, \textbf{\{Major;Minor\} AxisLength},\\
                \textbf{Perimeter}, \textbf{Mean \{Inside;Outside\} BoundaryIntensity}, \\
                \textbf{IntegratedIntensity},  \textbf{NormalizedBoundarySaliency}, \\
                Normalized \{Inside;\textbf{Outside}\} BoundaryIntensity, \\
                Eccentricity, Orientation, Intensity \{Mean;Range;Deviation\},\\
                Solidity, Circularity, EllipticalDeviation, BoundarySaliency, \\
                \{Inside;Outside\} BoundaryIntensity \{Range;Deviation\}
            \end{tabular} \\
\hline   
Voronoi features & 12 & 
            \begin{tabular}{@{}l@{}} \{Area;Perimeter;Chord\} \{Mean;Deviation;Disorder\},\\
                \{Area;Perimeter;Chord\}  Ratio of Minimum to Maximum
            \end{tabular} \\
\hline
\end{tabular}
\end{table}

\subsection{Supercell via Local Graph}
\begin{figure}
\includegraphics[width=\textwidth]{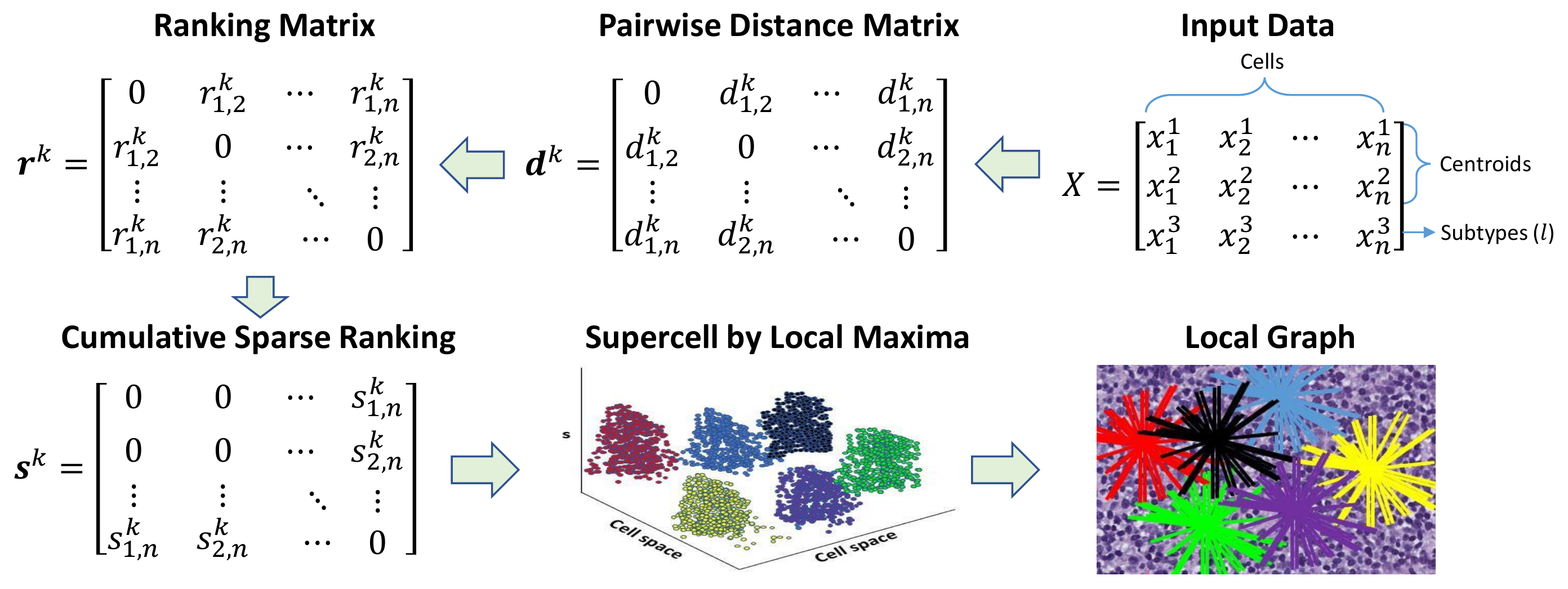}
\caption{The procedures of local graph construction via the ROC algorithm.}
\label{fig:local_graph_construction}
\end{figure}

In the local cell graph construction, the cell centroids together with the obtained cell subtypes information are used to identify cell clusters $\{\bm{C}\}$, where the Ranking Operation-based Clustering (ROC) algorithm applied (detailed in Fig.~\ref{fig:local_graph_construction})~\cite{gu2019distance}. The ROC algorithm is insensitive to the scale of features and type of distance metric used which makes it well-fitted for cell cluster identification. Given a set of $n$ cells, their feature vectors represented as $\bm{X}=\{\bm{x_i}\}_{i=1}^{n}$ where $\bm{x_i}\in\mathbb{R}^3$, the ROC algorithm first computes the pairwise distances $\{\bm{d}^{k}\}_{k=1}^{3}$ from $\bm{X}$. It then computes ranking matrices $\{\bm{r}^{k}\}_{k=1}^{3}$ containing the  ranking indices of the elements of $\{\bm{d}^{k}\}_{k=1}^{3}$. The cumulative sparse ranking of each cell feature $\{\bm{s}^{k}\}_{k=1}^{3}$ is then computed and all the local maxima $\{\bm{m}\}$ of the cumulative sparse ranking are identified and used to group the cells into clusters $\{\bm{C}\}$. 

\subsection{Cell Community Identification by Clustering of Supercells}
\begin{figure}[hbt!]
\includegraphics[width=\textwidth]{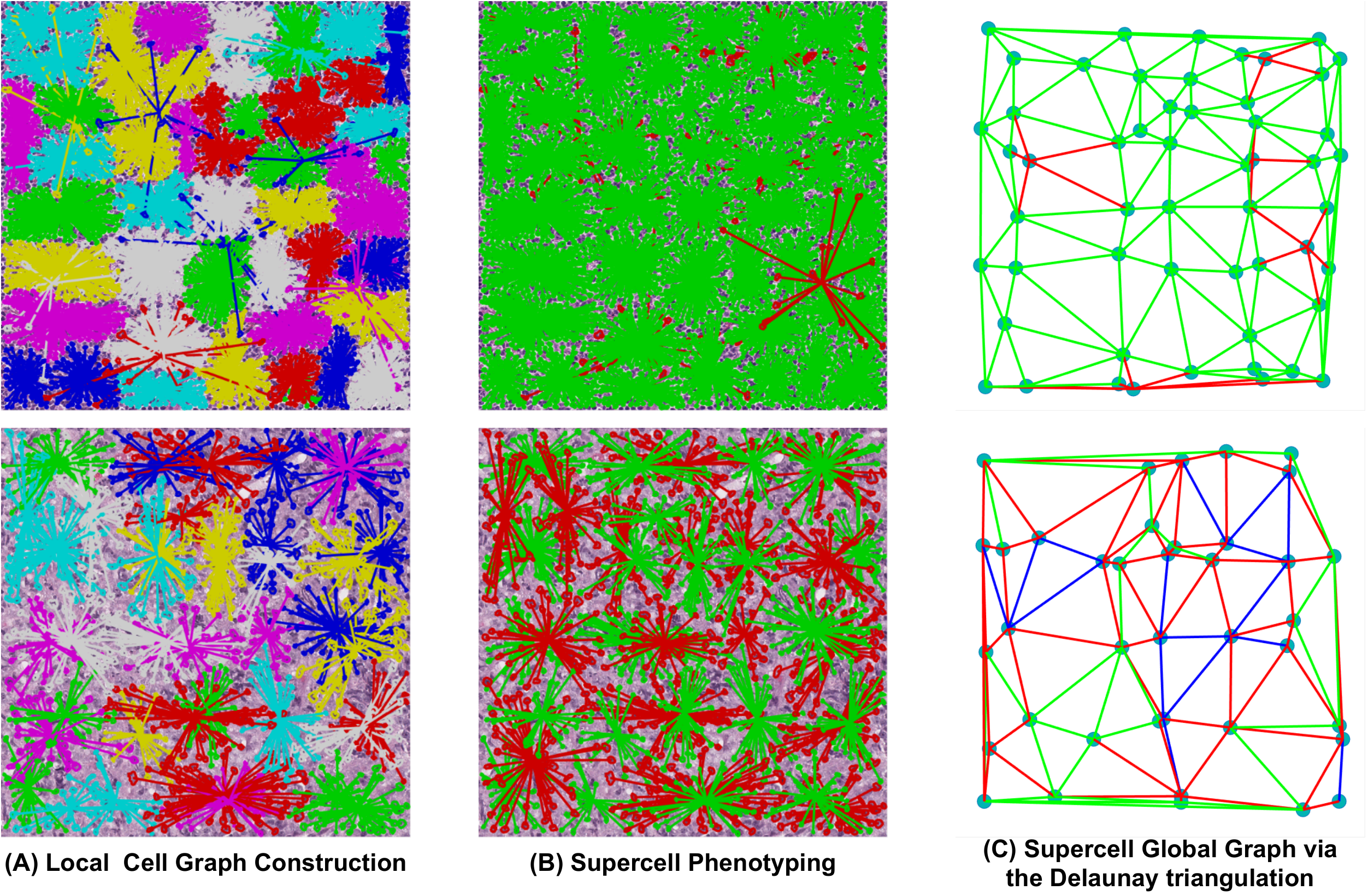}
\caption{Exemplar intermediate results of the proposed algorithm. Column (A) shows the constructed local graphs (Supercells). Column (B) shows the supercell phenotyping results. Column (C) shows the built global graphs via the Delaunay triangulation, in which different edge colors represent different supercell interaction patterns.}
\label{fig:process_flow}
\end{figure}

We take the convex hull-covered polygon as each supercell’s boundary. Unlike the individual cell, neighboring supercells often have overlapped regions, where the different cell classes interaction and confrontation occur (Fig.~\ref{fig:process_flow}(A)). To discover the supercell groups (we term as cell community), we perform unsupervised phenotyping at supercell level. We characterize an individual supercell by looking into: 1) the encompassed cell features (i.e., compute mean feature values of individual cells) as well as 2) the cellular spatial orchestration and interaction (i.e., build a Voronoi diagram within supercell and extract related features). With these features (n=22, bolded cell features, and all Voronoi features in Table 1), we implement clustering to identify supercell subtypes (identical to cellular level clustering in Section 2.1). As previously stated, t-SNE is first adopted to reduce the feature space and embed supercells. Then we use spectral clustering to define similar intra- and interpatient supercells (i.e., cell community). For the evaluated lymphocytic images, two types of cell communities are identified. Supercell phenotyping outcomes can be visualized from the two demo examples displayed in Fig.~\ref{fig:process_flow}(B).

\subsection{Global Supercell Graph Construction and Feature Extraction}
We build global graphs to characterize interactions among labeled supercells inside each image. Two global graphs, including the Delaunay triangulation and the Voronoi diagram, are constructed based on the supercell and cell community to complement each other. In the Delaunay scheme, the area is tessellated into triangular cells, which nodes are centroids of supercells. While in the Voronoi scheme, the area is split into sets of convex and adjacent polygonal cells, in which anchors are supercell centroids. Fig.~\ref{fig:process_flow}(C) shows two demos of the built global graph via the Delaunay triangulation. We extract features from the Delaunay graph through characterizing node-to-node interaction (i.e., edge) patterns. Assume there is $k$ different type of nodes (inherit from supercell labels), the overall number of edge types is $k(k+1)/2$. We accumulate each edge type’s counts and propose the percentages of different edge types as the global graph features. As for the global Voronoi graph, 12 features, as indicated in Table~\ref{tab:cell_features}, are extracted. We combine the features from both the Delaunay graph and the Voronoi diagram and test their performance in the diagnosis of different hematological malignancy subtypes (CLL vs aCLL vs RT-DLBL).

\section{Experiments and Results}
\subsection{Dataset Description}
Digitized clinical tumor tissue slides collected at The University of Texas MD Anderson Cancer Center (UTMDACC) were used a a study dataset. The study was approved by the UTMDACC Review Board and conducted following the Declaration of Helsinki. The glass slides were scanned using an Aperio AT2 scanner at 20x optical magnification, with each pixel equaling 0.5 \textmu m. We chose the best quality slide for each patient and cropped one image with a fixed size of $1,000 \times 1, 000$ pixels from level 0 of each pyramidal formatted slide. The image was cropped by a board-certified pathologist with subspecialty expertise who annotated the region of interest (ROI)’s central part. Each patient only has one sample image in this design to facilitate the k-fold cross-validation analysis. After excluding slides with server artifacts, we were left with 20 images of each diagnosis to maintain the balance among CLL, aCLL, and RT-DLBL.

\subsection{Implementation Details}
The implementation includes four aspects: cell phenotyping, local graph construction, supercell phenotyping, and image classification. Since each image with the size $1,000 \times 1, 000$ pixels contains thousands of cells, we randomly select 3,000 cells from each disease and pool 9,000 cells from all patients together to conduct the cell clustering. To propagate cell clustering labels to unsampled ones, we train the cell classifier with an ensemble of learners aggregated with the bagging method. In the local graph construction, there are two parameters to set for the ROC algorithm. We set the sparse ranking matrices’ sparsity as 0.9 and the number of neighbors for local maxima identification as 10. The total number of supercells generated at the population level is 2,794, and we use all of them for supercell phenotyping, which employs the same clustering and classification manners as in cell phenotyping. As for the image classification, we conduct repeated 5-fold cross-validation with the Support Vector Machine (SVM). We mainly employ two metrics to evaluate the performance, including accuracy and area under the curve (AUC) for each diagnosis with the one-vs-rest strategy. We report the mean values of these metrics with 100 times randomization. The code is released at \url{https://github.com/WuLabMDA/HierarchicalGraphModeling}.

\subsection{Results and Discussions}
We compare the proposed method with three cell-level graph-based algorithms, including the Global Cell Graph (GCG)~\cite{shin2015quantitative}, Local Cell Graph (LCG)~\cite{lewis2014quantitative}, and FLocK~\cite{lu2021feature}. The GCG is built using the Voronoi diagram; The LCG is constructed using the mean shift algorithm~\cite{tuzel2009kernel} utilizing the cell centroids; The FLocK is built with mean shift utilizing centroids as well as area and intensity of the cell. To have a fair comparison, we try our best to adjust the hyper-parameters used in these three compared methods to report their best results. The comparison results are shown in Table~\ref{tab:cmp_results}. Our proposed algorithm shows superior performance on these evaluated metrics among all the compared methods. The proposed method obtains the mean accuracy of 0.703 under 5-fold cross-validation with a 100 times randomization scheme for the three category disease diagnoses. Using only the local or the global graph fails to separate CLL from the other clinically relevant subtypes. By contrast, combining the local and global graph can improve the performance by a significant margin with a 0.4 increase in accuracy on the CLL. Further investigations find that the newly identified two cell communities have significantly different distributions between CLL and aCLL$+$RT (p-value$=$3.54e-08).

The two local graph methods (LCG and FLock) show better performance than the GCG. These results show that the local graph can better mine the cellular patterns than the global graph. Similar to the local graph methods, our method also firstly builds the local graph. Unlike them, LCG only utilizes the cell’s centroid information, and FLocK uses the centroids with some cell features. While the proposed method considers the cell type when building the local graph. Moreover, we construct another layer of cell community detection for the global graph on top of the created local graph to further abstract the cellular features. We assume that this two-layer graph algorithm is better at extracting the multi-scale (both local and global) cellular interactions and intratumoral heterogeneity, thus surpass the compared methods. To the best of our knowledge, this study is among the first to hybridize local and global graph methods to study cellular patterns. Hereby, we hypothesize that the proposed hybrid design can overcome the limitation inherent in the adoption of the global or local graph approaches solely to more meaningfully profile tumor composition and intratumoral heterogeneity. Indeed, the global approach misses the cellular level details while the local graph ignores the high-level interaction patterns between communities.

\begin{table}[ht]
\caption{Performance of the three compared graph theory based methods and the proposed method}
\centering 
\setlength{\tabcolsep}{6pt} 
\begin{tabular}{c c c c c} 
\hline
Method & Accuracy & AUC (CLL) & AUC (aCLL) & AUC (RT-DLBL) \\
\hline
GCG~\cite{shin2015quantitative}         & $0.436\pm0.037$ & $0.421\pm0.054$ & $0.730\pm0.027$ & $0.770\pm0.023$  \\
LCG~\cite{lewis2014quantitative}         & $0.471\pm0.042$ & $0.555\pm0.049$ & $0.669\pm0.050$ & $0.763\pm0.032$ \\
FLocK~\cite{lu2021feature}       & $0.601\pm0.045$ & $0.545\pm0.054$ & $\bm{0.816}\pm0.025$ & $0.847\pm0.022$  \\
Proposed    & $\bm{0.703}\pm0.030$ & $\bm{0.915}\pm0.009$ & $0.724\pm0.033$  & $\bm{0.866}\pm0.028$ \\
\hline 
\end{tabular}
\label{tab:cmp_results}
\end{table}

We take advantage of the clustering algorithm to attain the type of both cell and supercell (cell community) in hematolymphoid cancers, given that their cellular components in TME are unclear and that is distinct from solid tumors with well-studied TME. We attach great importance to cell type information and hypothesize that cells belonging to different types and communities play distinct functions. The unsupervised manner would also save tremendous workload for obtaining medical experts cell labeling and shed light on uncovering new insight into cell subtypes and communities.

We propose a novel supercell concept in this study. The supercell is anticipated to be an excellent bridge to connect the neighboring cell interaction (local graph) and cell community interaction (global graph). Here, we just scratch the surface of supercells properties, through merely exploiting supercells central coordinates to build global graphs. Indeed, each supercell contains rich information of the local TME, and their overlapped region may manifest interplay mechanisms of differing tumor subclones or habitats. We anticipate that future studies will help derive more biological insight based on the supercell context. Our approach in this analysis uses manually designed features from the supercell-based global graph for diagnosis due to limited image samples. However, future approaches should permit taking each supercell as the node to use the graph convolution network (GCN) to further improve diagnostic accuracy when sufficient images are available.

\section{Conclusion}
We present a hierarchical cell phenotyping and graph modeling algorithm to profile the spatial patterns of neoplastic lymphoid cells distributed within the tumor microenvironment. Using this global graph’s manual-crafted features, we achieve promising results based on clinical tumor images from various types of hematolymphoid tumors of different grades. We anticipate that the proposed cell phenotyping and supercell design can adapt to a broader category of cancers. 

%
\bibliographystyle{splncs04}
\bibliography{references}

\end{document}